\documentclass[preprint,3p,twocolumn]{elsarticle}

\usepackage{amssymb}
\usepackage{amsthm}
\usepackage{amsmath}
\usepackage{graphicx}
\usepackage{caption}
\usepackage{subcaption}
\usepackage{xcolor}
\usepackage{bm}
\usepackage{xurl}
\usepackage{caption}
\usepackage{float}
\usepackage{hyperref}

\begin{document}

\begin{frontmatter}

\title{Nuclear elastic scattering of protons below 250~MeV in FLUKA v4-4.0 and its role in single-event-upset production in electronics}

\author[inst1,inst2,inst3]{Alexandra-Gabriela \c{S}erban\corref{cor1}}
\ead{fluka.team@cern.ch}
\cortext[cor1]{Corresponding author}

\affiliation[inst1]{organization={European Organization for Nuclear Research},
    addressline={Esplanade des Particules 1},
    city={1211 Geneva 23},
    country={Switzerland}}

\affiliation[inst2]{organization={Faculty of Physics, University of Bucharest},
    addressline={405 Atomistilor},
    city={077125 Bucharest-Magurele},
    country={Romania}}
    
\affiliation[inst3]{organization={“Horia Hulubei” National Institute of Physics and Nuclear Engineering},
    addressline={30 Reactorului},
    city={077125 Bucharest-Magurele},
    country={Romania}}
    
\author[inst1]{Andrea Coronetti}
\author[inst1]{Rub\'{e}n Garc\'{i}a Al\'{i}a}
\author[inst1]{Francesc Salvat Pujol}

\author{\\on behalf of the FLUKA.CERN Collaboration}

\begin{abstract}

FLUKA is among the general-purpose codes for the Monte Carlo simulation of radiation transport that are routinely employed to estimate the production of single-event-upsets (SEUs) in commercial static random access memories (SRAMs) exposed to radiation. Earlier studies concerning the production of SEUs in commercial SRAMs under proton irradiation revealed very good agreement between experimental measurements and FLUKA estimates of the SEU production cross section for proton energies above 20-30~MeV. However, at lower proton energies, where the cross section for SEU production in such low-critical-charge components increases drastically, a FLUKA underestimation of up to two orders of magnitude was observed. Preliminary analyses indicated that this underestimation was in great measure due to the lack of nuclear elastic scattering of protons below 10~MeV in FLUKA up to version 4-3.4. To overcome this limitation, a new model for the nuclear elastic scattering of protons has been developed, combining partial-wave analyses and experimental angular distributions. This newly developed model has been included in FLUKA v4-4.0, and leads to an order-of-magnitude improvement in the agreement between FLUKA and experimental cross sections for the production of SEUs in SRAMs  under proton irradiation in the 1-10~MeV energy domain.

\end{abstract}


\end{frontmatter}

\section{Introduction}
\label{sec:intro}

Electronic systems such as those used in space missions, avionics,
and particle accelerator facilities may be damaged as a result of
exposure to radiation fields~\cite{SEP,avionics,huhtinen}. Electronic
components are sensitive to both cumulative radiation damage and
stochastic single-event effects (SEEs), which significantly disrupt
their operation. The radiation environment relevant to SEE production
in the context of particle accelerators is typically characterized
relying on the fluence of hadrons with energies higher than 20~MeV~\cite{ruben2017}. This
approximation assumes that the production of SEEs is mostly governed by
the energy deposition of fragments and recoiling residual nuclei from nuclear 
reactions of high-energy hadrons, thus neglecting contributions 
from direct ionization by hadrons, as well as from their elastic scattering on the
electronically screened target nuclei. However, recent studies have
shown that the production of SEEs, and particularly of
single-event-upsets (SEUs), induced by protons below 20~MeV is indeed
dominated by direct ionization~\cite{sierawski,cannon,caron} and elastic
scattering~\cite{coronetti,coronetti_paper,akkerman,zhenyu,caron}.

The role played by various radiation-matter interaction mechanisms in
the production of SEUs can be quantitatively assessed through Monte
Carlo simulations of particle transport, employing general purpose codes,
\textit{e.g.}\ FLUKA~\cite{flukaweb,batt,frontiers}, readily accessible
through \url{https://fluka.cern} for non-commercial purposes. FLUKA
simulates the coupled hadronic and electromagnetic showers set up in
complex material geometries by more than 60 particle species, with
energies from the keV up to the PeV domain, with neutrons exceptionally
tracked down to thermal energies. Thus, FLUKA covers a broad range of
applications, from particle accelerator design and operation, to
radiation protection aspects, medical applications, cosmic ray physics, to
name a few, in the interest of a community of nearly 4000 users. FLUKA
is among the simulation tools employed by the Radiation to Electronics
(R2E) team at CERN~\cite{lerner,cecchetto,ruben_thesis}, which ensures the successful operation of the
accelerator infrastructure taking into account the effects of radiation
exposure on electronic components and systems.  

\begin{figure}
\centering
\includegraphics[width=\columnwidth]{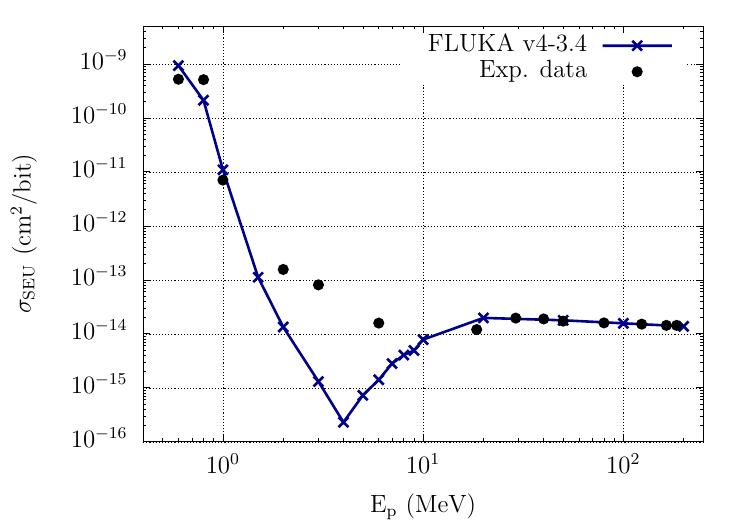}
\caption{Comparison between experimental SEU production cross section (black dots; uncertainties are smaller than the symbol size)~\cite{coronetti,coronetti_paper} and FLUKA v4-3.4 predictions (dark-blue crosses with solid line to guide the eye) induced by low-energy protons in the ISSI SRAM. See text for detailed discussion.}
\label{fig:seu_intro}
\end{figure}

In a recent R2E study~\cite{coronetti,coronetti_paper}, the production of SEUs in an
ISSI SRAM~\cite{issi} under proton irradiation was assessed. Figure
\ref{fig:seu_intro} displays the cross section for SEU production,
$\sigma_\text{SEU}$, in this device as a function of the proton energy.
Black dots display the experimental measurements, for which the uncertainty is smaller than the symbol size, while the dark-blue
crosses (connected with a solid line to guide the eye) represent the
FLUKA v4-3.4 prediction. For proton energies above 20-30~MeV,
where SEU production is driven by nuclear reactions~\cite{caron}, remarkable
agreement was obtained. However, in the 1-10~MeV range, where
$\sigma_\text{SEU}$ drastically increases towards lower energies,
FLUKA exhibits an underestimation of up to two orders of magnitude.
Preliminary analyses suggested that in the 1-10~MeV
range, proton nuclear elastic scattering significantly contributes to
SEU production in the ISSI SRAM~\cite{coronetti}. However, this interaction mechanism was
not available for protons below 10~MeV as of FLUKA v4-3.4. Furthermore,
above 10~MeV, a too simplistic account of large-scattering-angle
deflections was provided, often over- or under-estimating their
importance. These deflections, however, contribute significantly to the
production of SEUs, even at energies of up to $\mathcal{O}(100)$~MeV~\cite{coronetti}. To overcome these limitations, a new model for the
nuclear elastic scattering of protons from Coulomb barrier up to 250~MeV
has been developed and included in FLUKA v4-4.0. 

This work is structured as follows. In Section~\ref{sec:ranft} the
drawbacks of the FLUKA v4-3.4 model for proton nuclear elastic
scattering are briefly outlined. In Section~\ref{sec:newmodel} the
FLUKA v4-4.0 model for the nuclear elastic scattering of protons up
to 250~MeV is presented, highlighting its advantages over the preceding
model. The performance of this newly implemented model is assessed in
Section~\ref{sec:R2E}, comparing FLUKA v4-4.0 estimates with 
experimental measurements of the SEU production cross sections induced 
by low-energy protons in the aforementioned commercial ISSI SRAM. Finally, Section~\ref{conclusions} provides
both a summary of this work, and an outlook on future works detailing
recent benchmarking efforts to assess and document the good performances of this new model
in further SEU-production scenarios, as well as in energy deposition 
by proton beams in water phantoms for medical physics applications.

\section{Proton nuclear elastic scattering as of FLUKA v4-3.4}
\label{sec:ranft}

FLUKA is a condensed-history Monte Carlo code relying on the Moli\`ere
multiple Coulomb scattering (MCS) theory for an aggregate description of
the effect of multiple elastic collisions of charged particles on the
electrostatic potential of atoms along macroscopic particle steps in
matter~\cite{ferrari}. In this condensed-history approach, a screened
Rutherford differential cross section (DXS) is assumed; finite-size
effects and spin-relativistic corrections are treated by means of form
factors~\cite{ferrari,tsai}. While this approach is effective for the elastic scattering of
leptons, it is insufficient for that of charged hadrons since, in addition to
the Coulomb force, they are subject to the strong nuclear force.
Unfortunately, the Moli\`ere MCS theory does not allow for an easy
extension to account for the latter. Thus, the effect of the nuclear
force on the elastic scattering of charged hadrons must be accounted for
by a separate interaction mechanism, henceforth called nuclear elastic
scattering. Formal difficulties are encountered when treating nuclear
and Coulomb elastic scattering as separate interaction mechanisms
(elucidated in Section~\ref{sec:disentangling}), especially near Coulomb
barrier. For this reason, the proton nuclear elastic scattering model of
FLUKA v4-3.4 is not active below 10~MeV.  

\begin{figure}[ht]
  \centering
  \begin{subfigure}[b]{\columnwidth}
    \centering
    \includegraphics[width=\columnwidth]{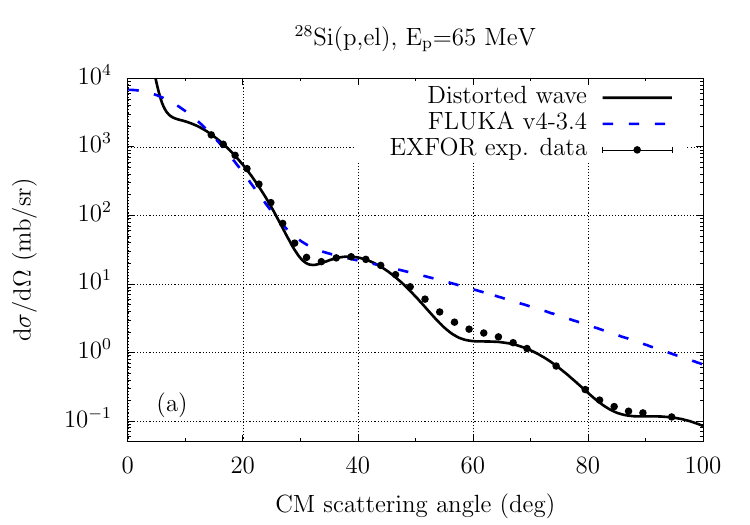}
  \end{subfigure}
  \hfill
  \begin{subfigure}[b]{\columnwidth}
    \centering
    \includegraphics[width=\columnwidth]{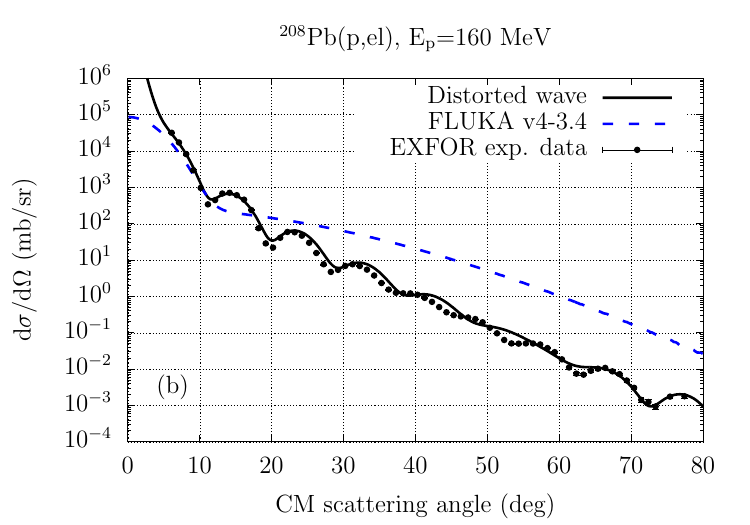}
  \end{subfigure}
  \caption{(a) Differential cross section for 65~MeV protons elastically scattering on $^{28}$Si. (b) Same as (a) for 160~MeV protons on $^{208}$Pb. See text for detailed discussion.}
  \label{fig:ranft_vs_exp_vs_dw}
\end{figure}

As of FLUKA v4-3.4, the DXS for the nuclear elastic scattering of
protons is based on an effective bimodal description attempting to
capture on the one hand the forward-scattering peak and on the other
hand the large-scattering-angle domain~\cite{ranft}. Figure~\ref{fig:ranft_vs_exp_vs_dw}a displays the DXS for the elastic
scattering of 65~MeV protons on $^{28}$Si as a function of the
scattering angle in the center-of-mass frame (CM). Experimental data~\cite{exfor,zerkin} are represented by the black dots (uncertainties are
smaller than the symbol size), while the DXS for nuclear elastic
scattering sampled from FLUKA v4-3.4 is displayed by the dashed blue
curve (statistical uncertainties omitted for clarity);  the black solid
curves are discussed in Section~\ref{sec:partial-wave}. The dominant
forward scattering feature extending up to $\sim$30~deg is indeed
reasonably reproduced. However, at large scattering angles there is an
order-of-magnitude discrepancy which is even more accentuated for
heavier target nuclei and higher proton energies, see \textit{e.g.}\
Fig.~\ref{fig:ranft_vs_exp_vs_dw}b for 160~MeV protons elastically
scattering on $^{208}$Pb. 

Up to FLUKA v4-3.4, an effective integrated cross section for proton
nuclear elastic scattering was obtained based on parametrizations of
the total and nuclear reaction cross sections of neutrons~\cite{angeli-csikai}. This effective integrated cross section
typically provides the DXS with the correct intensity as far as the
forward-scattering peak is concerned, see \textit{e.g.} Fig.~\ref{fig:ranft_vs_exp_vs_dw}. At large
scattering angles, instead, neither the intensity nor the structure of
minima and maxima of the DXS are correctly reproduced. Unfortunately, as discussed in Section~\ref{sec:R2E}, elastic collisions with large 
scattering angle strongly contribute to SEU
production. Therefore, particular effort has been made in the modelling
work discussed in the next section to characterize such collisions. 

\section{New model for proton nuclear elastic scattering in FLUKA v4-4.0}
\label{sec:newmodel}

To overcome the limitations outlined in the foregoing sections, a new
model for the nuclear elastic scattering of protons has been developed
and included in FLUKA v4-4.0. The optical potential model (OPM) of
Koning and Delaroche~\cite{koning} has been employed to effectively
describe the interaction of protons of up to 250~MeV with target nuclei
with  mass number $A\geq20$, and to evaluate a database of DXSs by means
of a partial-wave analysis (see Section~\ref{sec:partial-wave}). To
minimize memory requirements, an effective parametrized expression
depending on 7 parameters has been fitted onto the calculated DXSs
database (see Section~\ref{sec:parametrization}). For lighter target
nuclei, where OPMs are scarcer, the parametrized expression has been directly fitted onto available experimental angular distributions (see
Section~\ref{sec:light_targets}). An effective integrated cross section
for proton nuclear elastic scattering has been obtained by numerical
integration of the aforementioned parametrized expression. Finally, an
algorithm to numerically sample nuclear elastic scattering events has
been implemented (see Section~\ref{sec:integrated_xs}). The error
incurred when treating Coulomb and nuclear elastic scattering as
separate interaction mechanisms has been assessed in
Section~\ref{sec:disentangling}.

\subsection{Partial-wave analysis}
\label{sec:partial-wave}

The DXS for the elastic scattering of non-relativistic spin-1/2 particles on a central potential as a function of the polar and azimuthal CM scattering angles $\theta$ and $\varphi$ is given by
\begin{equation}
    \frac{\mathrm{d} \sigma}{\mathrm{d} \hat{\boldsymbol{\Omega}}}
     =\big|f(\hat{\boldsymbol{\Omega}})\big|^2
     +\big|g(\hat{\boldsymbol{\Omega}})\big|^2,
    \label{eq:dxspw}
\end{equation}
where $\hat{\bm{\Omega}}=(\theta,\varphi)$, while
\begin{equation}
\begin{aligned}
    f(\hat{\bm{\Omega}}) = 
        f_{\mathrm{C}}(\theta)
        & +
        \frac{1}{2 \hspace{0.025cm} \mathrm{i} k} \sum_{\ell=0}^{\infty} P_{\ell}(\cos\theta) \hspace{0.05cm} \textrm{e} ^{\textrm{i}2\Delta_\ell}\\
        & \times \big[ (\ell+1) \hspace{0.05cm} \mathrm{e}^{\mathrm{i} 2 {\delta}_{\ell, \ell+1 / 2}} \\
        & + \ell \hspace{0.05cm} \mathrm{e}^{\mathrm{i} 2 {\delta}_{\ell, \ell-1 / 2}}-(2 \ell+1) \big] 
\end{aligned}
   \label{eq:direct}
\end{equation}
and
\begin{equation}
   \begin{aligned}
     g(\hat{\bm{\Omega}}) = 
    \frac{1}{2 \hspace{0.025cm} \mathrm{i} k} \mathrm{e}^{i \varphi} 
    & \sum_{\ell=1}^{\infty} P_{\ell}^1(\cos \theta) \\
    & \times \left(\mathrm{e}^{\mathrm{i} 2 {\delta}_{\ell, \ell+1 / 2}} - \mathrm{e}^{\mathrm{i} 2 {\delta}_{\ell, \ell-1 / 2}}\right)
   \end{aligned}
   \label{eq:spin_flip}
\end{equation}
are the direct and the spin flip scattering amplitudes, respectively.
The CM wavevector is denoted by $k$, $P_\ell$ are the Legendre polynomials, and $P_\ell^m$ are the associated Legendre polynomials. The Coulomb scattering amplitude on a point nucleus is denoted by
$f_\mathrm{C}(\theta)$, while $\Delta_\ell$ and $\delta_{\ell,j}$ are the Coulomb and the inner phase shifts, respectively. While the former are analytical, the latter have been obtained in this work by numerically solving the radial Schr\"odinger equation for protons in the aforementioned OPM using the RADIAL subroutine package~\cite{radial}.

Defining
\begin{equation}
\begin{aligned}
    f_\mathrm{N}(\theta) & = \frac{1}{2 \hspace{0.025cm} \mathrm{i} k} 
    \sum_{\ell=0}^{\infty} P_{\ell}(\cos\theta) \hspace{0.05cm} \textrm{e} ^{\textrm{i}2\Delta_\ell}\\
    & \times \big[ (\ell+1) \hspace{0.05cm} \mathrm{e}^{\mathrm{i} 2 {\delta}_{\ell, \ell+1 / 2}} \\
    & + \ell \hspace{0.05cm} \mathrm{e}^{\mathrm{i} 2 {\delta}_{\ell, \ell-1 / 2}}-(2 \ell+1) \big]
\end{aligned}
\end{equation}
and inserting Eqs.~\eqref{eq:direct} and \eqref{eq:spin_flip} in Eq.~\eqref{eq:dxspw}, the DXS becomes
\begin{equation}
\begin{aligned}
    \frac{\mathrm{d}\sigma}{\mathrm{d}\hat{\boldsymbol{\Omega}}}
    &= \big|g(\hat{\boldsymbol{\Omega}})\big|^2 + \big|f_\mathrm{C}(\theta)\big|^2 + \big|f_\mathrm{N}(\theta)\big|^{2}+ \\
    & + 2 \hspace{0.05cm} \mathrm{Re}\left[f_\mathrm{C}^{*}(\theta)f_\mathrm{N}(\theta)\right].
\end{aligned}
\end{equation}
While the first three terms are positive, the remaining interference term between Coulomb and nuclear elastic scattering can be either positive or negative. This formally precludes treating nuclear and Coulomb elastic scattering as additive (separate) interaction mechanisms, especially at energies near Coulomb barrier, hence the lack of nuclear elastic scattering below 10~MeV up to FLUKA v4-3.4. Unfortunately, as mentioned in Section~\ref{sec:ranft}, the use of Moli\`ere MCS scheme in FLUKA necessarily implies a separate treatment of Coulomb and nuclear elastic scattering. In Section~\ref{sec:disentangling} the error incurred by this approach is discussed.  

\begin{figure}
\centering
\includegraphics[width=\columnwidth]{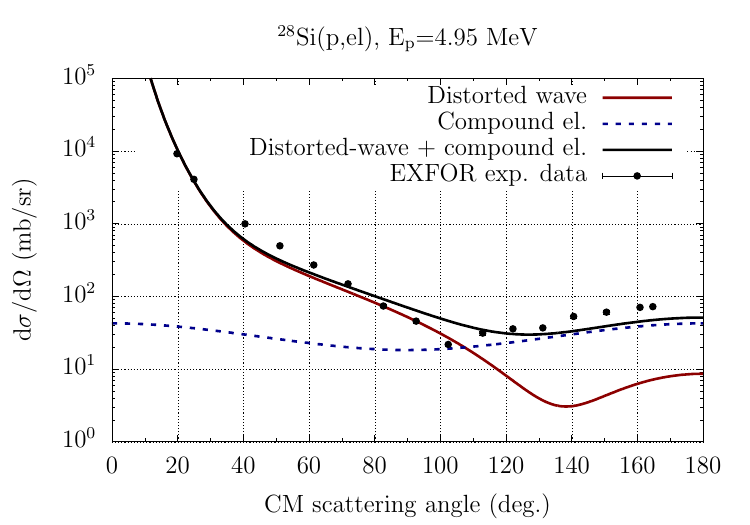}
\caption{Differential cross section for the elastic scattering of 4.95~MeV protons on $^{28}$Si with (black curve) and without (red curve) the contribution of the compound elastic scattering (dashed blue curve) compared with experimental DXSs (black dots; uncertainties are smaller than the symbol size)~\cite{exfor,zerkin}. See text for detailed discussion.}
\label{fig:compound}
\end{figure}

Relying on a dedicated implementation of the partial-wave scheme outlined above, a database of DXSs for the elastic scattering of protons has been evaluated on a grid of 14 target nuclei, from $^{20}$Ne to $^{238}$U, and on a grid of 37 proton energies, from Coulomb barrier up to 250~MeV. To confirm the soundness of the implemented partial-wave scheme, a systematic benchmark has been performed, wherein calculated DXSs have been compared with experimental DXSs~\cite{exfor,zerkin}. Overall, good agreement has been obtained\footnote{Also for isotopes slightly beyond the strict domain of applicability of the employed OPM, \textit{e.g.}\ $^{20}$Ne and $^{238}$U.}, as shown in Fig.~\ref{fig:ranft_vs_exp_vs_dw}, where the black solid curves represent the DXSs calculated with the partial-wave scheme adopted here. At most, deviations in the order of ~20-30\% are occasionally encountered in narrow angular domains, due to the use of a globally fitted OPM instead of a local fit.

Finally, following Koning and Delaroche~\cite{koning}, the calculated DXSs include an account of compound nuclear elastic scattering, based on Ref.~\cite{auerbach}
for target nuclei up to $^{40}$Ar and proton energies up to 15~MeV. This additional contribution significantly improves the agreement with experimental angular distributions in the large-scattering-angle domain at low energies, as shown in Fig.~\ref{fig:compound} for the elastic scattering of 4.95~MeV protons on $^{28}$Si.

\subsection{Parametrized differential cross section}
\label{sec:parametrization}

The database of tabulated DXSs outlined above could have been readily
adopted for sampling proton nuclear elastic scattering events in FLUKA.
However, a database evaluation in a sufficiently dense grid of energies,
target nuclei, and scattering angles would promptly lead to memory requirements in the
order of tens if not hundreds of megabytes. Dedicating such an amount of
memory from which a single interaction mechanism (nuclear elastic scattering) 
for a single particle species (protons) in a restricted energy range 
(from Coulomb barrier to 250~MeV) would benefit, has been discarded in the 
framework of a general-purpose multi-particle tracking code such as FLUKA.

Thus, an effective analytical DXS has been sought with sufficient flexibility
to reasonably reproduce the structure of maxima and minima of the actual
DXS. In a spirit similar to that of Refs.~\cite{ferrari1998,dremin,grichine}, the DXS for the elastic scattering
of a particle on a fully absorptive imaginary potential (in the so-called
black-disk limit~\cite{dremin}) has been recast as follows:
\begin{equation}
\begin{aligned}
\label{eq:paramexpr}
        \cfrac{\mathrm{d}\sigma_0}{\mathrm{d}\hat{\boldsymbol{\Omega}}}  
    = 
    \alpha \hspace{0.025cm} k^2 R^{4} 
    \Bigg[ 
        &\Bigg( 
            \cfrac{J_1(R \hspace{0.025cm} q \hspace{0.025cm}  \delta_1)}
                  {R \hspace{0.025cm} q} 
        \Bigg)^2 
        e^{-\beta_1 R \hspace{0.025cm} q}        
        \\
    & +\gamma J_0^2(R \hspace{0.025cm} q \hspace{0.025cm} \delta_0) 
        e^{-\beta_0 R \hspace{0.025cm} q} 
    \Bigg],
\end{aligned}
\end{equation}
where $J_1$ and $J_0$ are Bessel functions of the 1$^{\text{st}}$ kind, and $R=1.2 \hspace{0.05cm} A^{1/3}$ is the nuclear radius in fm which provides a built-in scaling with the target mass number $A$. Furthermore,
\begin{equation}
\label{eq:qcm}
    q = 2 \hspace{0.025cm} k \sin \bigg(\frac{\theta}{2}\bigg),
\end{equation}
is the CM wavevector transfer expressed as a function of the CM wavevector $k$ and the CM scattering angle $\theta$. The quantities $\alpha$, $\beta_{0,1}$, $\gamma$, $\delta_{0,1}$ are 6 dimensionless fit parameters.

\begin{figure}
\centering
\includegraphics[width=\columnwidth]{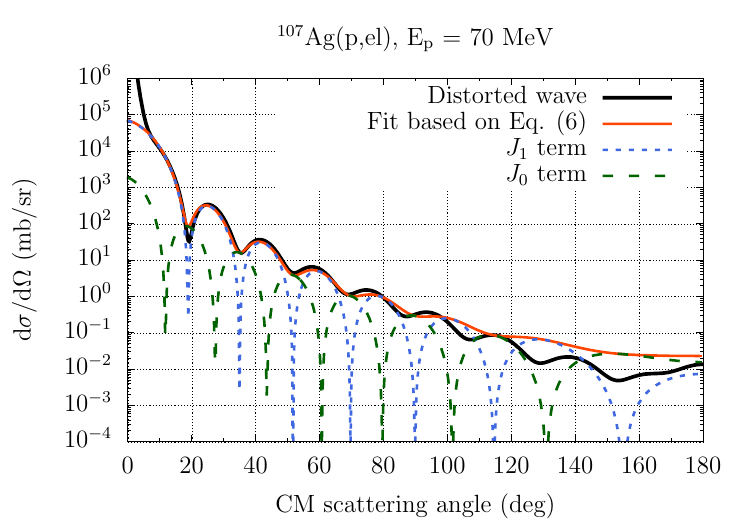}
\caption{Differential cross section for the elastic scattering of 70~MeV protons on $^{107}$Ag. See text for detailed discussion.}
\label{fig:j1j0}
\end{figure}

As an example, the thick black curve in Fig.~\ref{fig:j1j0} displays the DXS for the elastic scattering of 70~MeV protons on $^{107}$Ag calculated within the partial-wave approach discussed in the foregoing section, while the thin orange curve represents the DXS obtained by fitting parametrized expression \eqref{eq:paramexpr}, yielding $\alpha=6.141$, $\beta_0=0.296$, $\beta_1=0.369$, $\gamma=0.008$, and $\delta_0=\delta_1=1.105$. The drastic rise of the DXS as the scattering angle approaches 0~deg is instead a feature captured by Coulomb scattering, as discussed in Section~\ref{sec:disentangling}. The dashed green and blue curves in Fig.~\ref{fig:j1j0} show the contributions of the $J_1$ and $J_0$ terms of Eq.~\eqref{eq:paramexpr}. Their minima and maxima are in phase opposition, allowing to capture the structure of minima and maxima of the actual DXS with a certain degree of flexibility, as requested. The 6 fit parameters play different roles: $\alpha$ is a mere scaling factor; $\beta_{1}$ and $\beta_{0}$ adjust the slope of the DXS; $\delta_{1}$ and $\delta_{0}$ allow for flexibility in capturing the position of the minima and maxima of the DXS, while $\gamma$ drives the depth of the minima in the DXS. 

\begin{figure}
\centering
\includegraphics[width=\columnwidth]{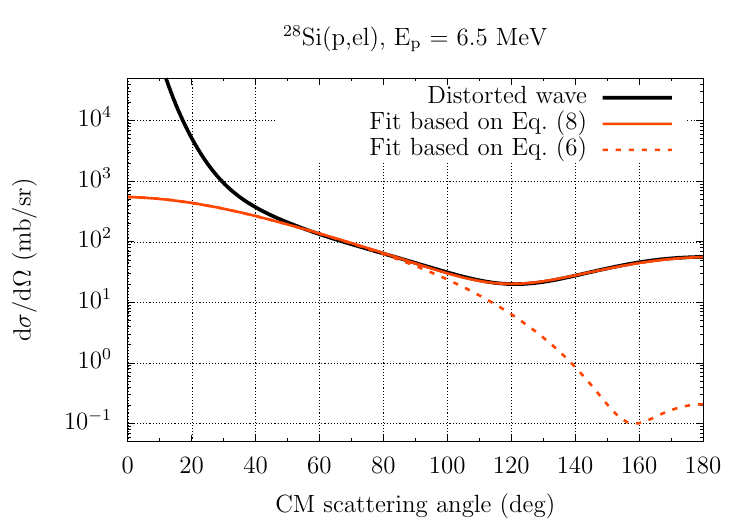}
\caption{Differential cross section for the elastic scattering of 6.5~MeV protons on $^{28}$Si. See text for detailed discussion.}
\label{fig:zeta}
\end{figure}

\begin{figure*}[ht]
\centering
\includegraphics[]{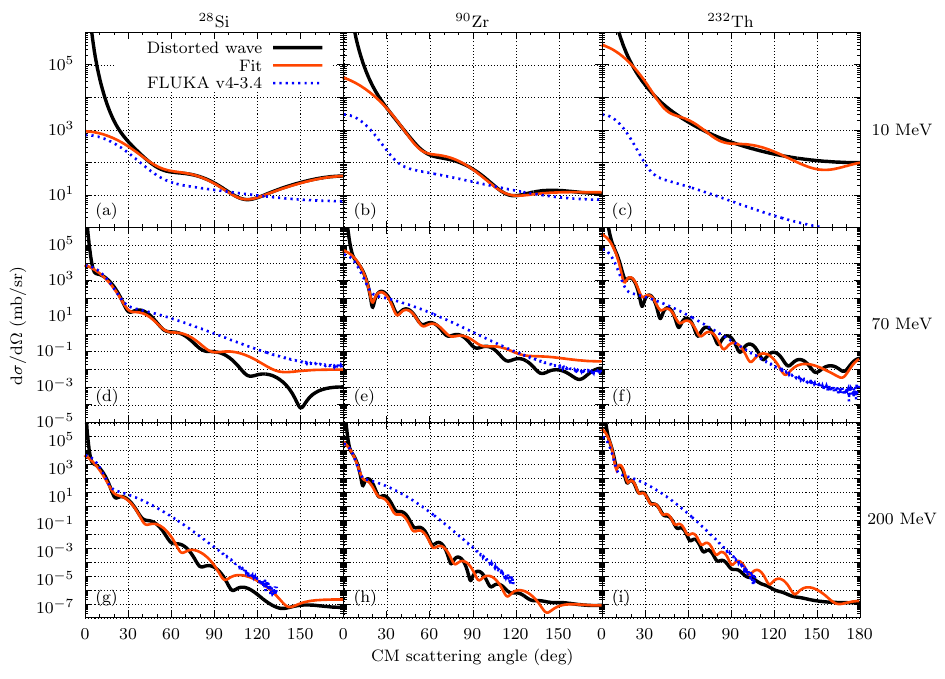}
\caption{Fitted parametrized expression (thin orange curves) compared with DXSs evaluated within the partial-wave approach (thick black curves) for protons of 10~MeV, 70~MeV, and 200~MeV (first, second, and third row, respectively) elastically scattering on $^{28}$Si, $^{90}$Zr, and $^{232}$Th (first, second, and third column, respectively), and with the FLUKA v4-3.4 sampled DXSs for nuclear elastic scattering (dashed blue curves). See text for detailed discussion.}
\label{fig:fits}
\end{figure*}

As shown in Fig.~\ref{fig:j1j0}, the proposed parameterized expression
\eqref{eq:paramexpr} is able to reproduce not only the main forward
scattering feature extending up to $\mathcal{O}(10)$~deg, but also a considerable amount of minima
and maxima at large scattering angles. Occasional difficulties are
however encountered when trying to capture accentuated backscattering
features. Therefore, it has been decided to provide Eq.~\eqref{eq:paramexpr}
with a further term, prefaced by an additional fit parameter $\zeta$,
\begin{align}
\label{eq:fullexpression}
  \frac{\mathrm{d}\sigma}{\mathrm{d}\hat{\boldsymbol{\Omega}}}
  =
  \frac{\mathrm{d}\sigma_0(\theta)}{\mathrm{d}\hat{\boldsymbol{\Omega}}}
  +
  \zeta
  \frac{\mathrm{d}\sigma_0(\pi-\theta)}{\mathrm{d}\hat{\boldsymbol{\Omega}}},
\end{align}
which facilitates the fit of large-angle scattering features. The effect of this additional term is highlighted in Fig.~\ref{fig:zeta} which displays in thick black curve the DXS calculated within the partial-wave scheme of Section~\ref{sec:partial-wave}, in dashed orange the DXS obtained by fitting Eq.~\eqref{eq:paramexpr}, missing the prominent backscattering feature, and in solid orange the DXS obtained by fitting Eq.~\eqref{eq:fullexpression}, yielding excellent agreement with the distorted-wave calculated DXS.

Finally, the adopted parametrized DXS \eqref{eq:fullexpression} depends
on 7 fit parameters ($\alpha,\beta_{0,1},\gamma,\delta_{0,1},\zeta$)
which have been fitted to the database of DXSs generated as described in the
previous section\footnote{See Section~\ref{sec:light_targets} for the
analogous treatment of light target nuclei.} by means of a dedicated least-squares
minimization. This effort has effectively reduced the memory requirements to a mere
tabulation of 7 parameters for 14 target nuclei and 37 tabular proton energies. 
Fit parameter values have been obtained with an uncertainty of $\pm 2.5\%$. A goodness-of-fit test has revealed that in $88.6\%$ of the
considered energy and target tuples, the obtained fit parameters can be
accepted with a confidence level of 5\%. 

Figure \ref{fig:fits} displays the DXS for the elastic scattering of protons on $^{28}$Si, $^{90}$Zr and $^{232}$Th (first, second, and third column, respectively) at 10~MeV, 70~MeV, and 200~MeV (first, second, and third row, respectively). The thick black curves have been evaluated with the partial-wave approach discussed in Section~\ref{sec:partial-wave}, while the thin orange curves have been obtained by fitting parametrized expression \eqref{eq:fullexpression}. Finally, the dashed blue curves have been obtained by sampling nuclear elastic scattering events from FLUKA v4-3.4 and scaling the resulting (unit-normalized) angular distributions by the integrated cross section detailed in Section~\ref{sec:ranft}. The resulting DXSs generally capture the forward scattering feature, but tend to provide a too coarse account of the large-scattering-angle domain, missing the rich structure of maxima and minima exhibited by the actual DXSs. Moreover, subfigure (c) exemplifies a case in which the FLUKA v4-3.4 integrated cross section for nuclear elastic scattering provides insufficient intensity (see also Section~\ref{sec:disentangling}). Instead, parametrized expression~\eqref{eq:fullexpression} captures not only the forward-scattering feature, but also a considerable number of minima and maxima at large scattering angles, especially at high energies and for large mass numbers, as shown by the thin orange curves in subfigures (f), (h) and (i). However, for low mass numbers, at localized energies in the few tens of MeV, the agreement is occasionally less optimal at large scattering angles (where, nevertheless, the DXSs have already dropped by several orders of magnitude), as shown in subfigures (d), (e) and (g). Furthermore, for intermediate and large mass numbers and proton energies near Coulomb barrier, mild wiggling of the fitted parametrized expression around the actual DXS is encountered, as seen in subfigure (c). Nonetheless, the parametrized expression fulfills its original purpose to provide a good description of the forward-scattering peak, as well as a fairly resolved account of the structure of minima and maxima at larger scattering angles, relying only on 7 fit parameters.

\begin{figure}
\centering
\includegraphics[width=\columnwidth]{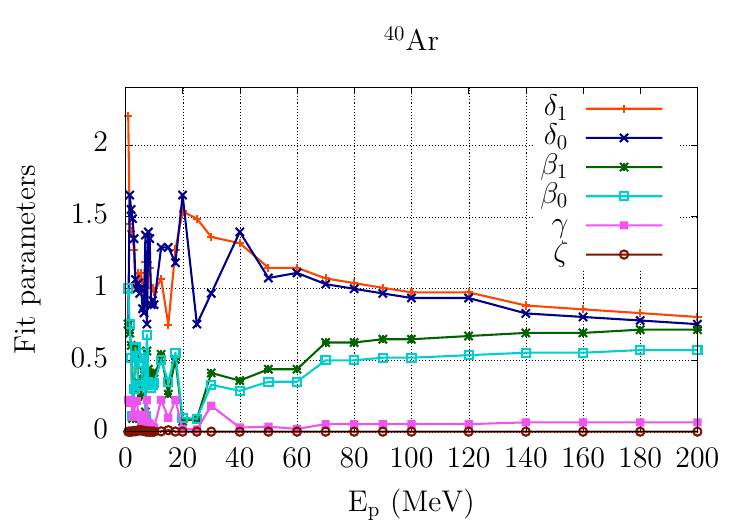}
\caption{Fit parameters as a function of proton energy for $^{40}$Ar. See text for detailed discussion.}
\label{fig:parameters}
\end{figure}

\begin{figure*}[ht]
\centering
\includegraphics[]{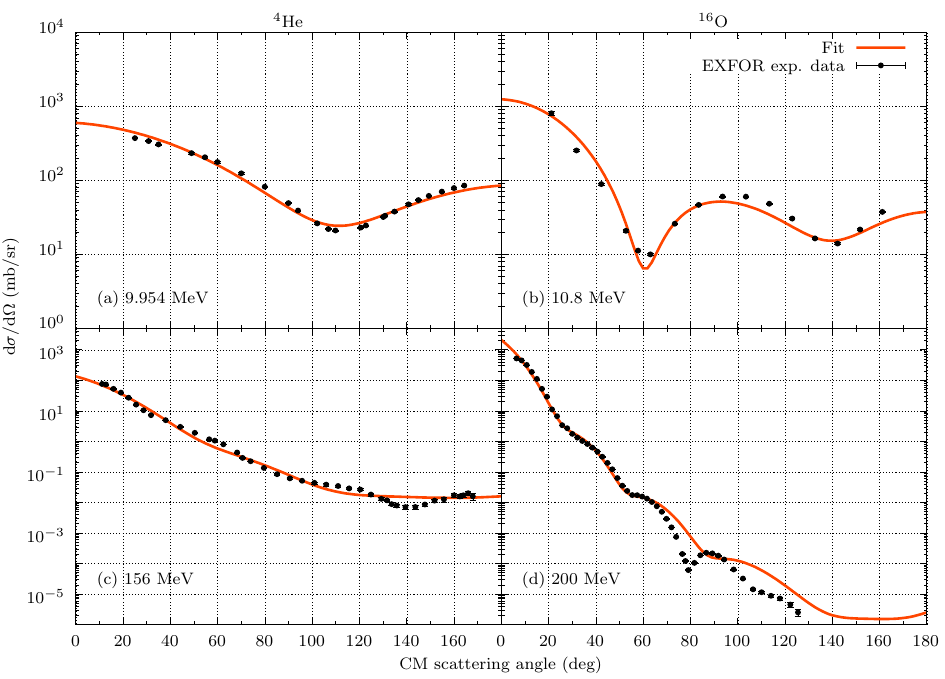}
\caption{Fitted parametrized expression (solid orange curves) compared with experimental DXSs (black dots; uncertainties are smaller than the symbol size)~\cite{exfor,zerkin} for protons of various energies on $^{4}$He (left) and $^{16}$O (right). See text for detailed discussion.}
\label{fig:lighttar}
\end{figure*}

Figure \ref{fig:parameters} displays the values of all relevant fit parameters ($\alpha$ is not needed for numerical sampling purposes) as a function of the proton energy for the nuclear elastic scattering of protons on $^{40}$Ar. At energies above a few tens of MeV, the energy dependence is smooth since the parametrized expression \eqref{eq:fullexpression} relies on the black-disk limit, \textit{i.e.}\ it works best at high proton energies. At lower energies, the fit parameters exhibit a less smooth behaviour. Nevertheless, their values do not significantly deviate from $\mathcal{O}(1)$. To minimize posterior interpolation errors at low energies, the energy grid is roughly logarithmic. 

\subsection{Nuclear elastic scattering on light target nuclei}
\label{sec:light_targets}

The Koning and Delaroche OPM is not strictly applicable to model the elastic scattering of protons on target nuclei with mass number $A<24$. Moreover, global OPMs for protons on light nuclei are not readily available. Thus, a complementary strategy has been adopted: the parametrized expression \eqref{eq:fullexpression} has been directly fitted on the available experimental DXSs~\cite{exfor,zerkin} for 13 target nuclei with mass numbers from $A=2$ to $A=16$, for proton energies below 250~MeV. Figure \ref{fig:lighttar} displays the good fit of parametrized expression \eqref{eq:fullexpression} (represented by the solid orange lines) on the experimental DXSs (in black dots; uncertainties are smaller than the symbol size) for the elastic scattering of protons of various energies on $^{4}$He (first column) and $^{16}$O (second column). The proton-proton nuclear elastic scattering model has not been altered with respect to FLUKA v4-3.4.

\subsection{Implementation in FLUKA v4-4.0}
\label{sec:integrated_xs}

A sampling scheme has been implemented in FLUKA v4-4.0 for the simulation of proton nuclear elastic scattering events from the fitted parametrized expression \eqref{eq:fullexpression}. Since the inverse sampling equation for this distribution does not have an analytical solution, a rejection sampling scheme has been adopted using the exponential terms in Eq.~\eqref{eq:fullexpression} as reasonable envelope functions and the remaining Bessel-function terms as acceptance weights. 

\begin{figure}
\centering
\includegraphics[width=\columnwidth]{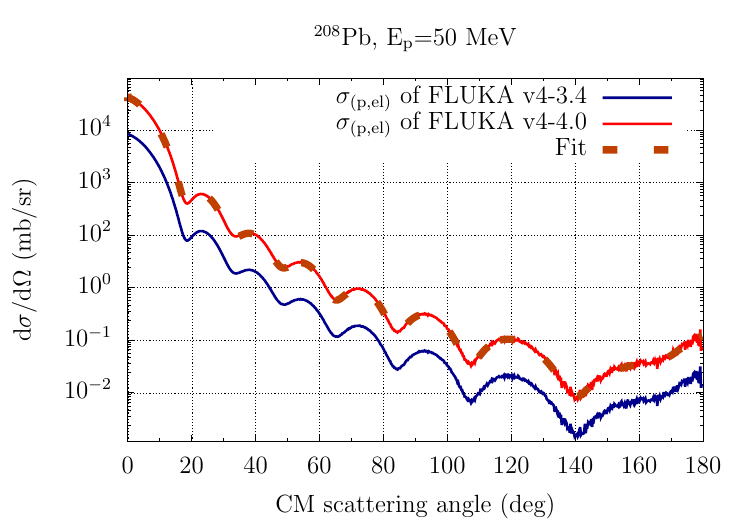}
\caption{DXS for the nuclear elastic scattering of 50~MeV protons on $^{208}$Pb when using the proton nuclear elastic scattering integrated cross section of FLUKA v4-3.4 (solid blue curve) vs.\ when using that of FLUKA v4-4.0 (solid red curve) vs.\ fitted parametrized expression \eqref{eq:fullexpression} (thick dashed dark-orange curve). See text for detailed discussion.}
\label{fig:xs}
\end{figure}

In Fig.~\ref{fig:xs}, the thick dashed dark-orange curve displays the
fitted DXS based on Eq.~\eqref{eq:fullexpression} for the nuclear
elastic scattering of 50~MeV protons on $^{208}$Pb, while the blue solid
curve represents the sampled DXS one would obtain relying on the
unaltered integrated cross section for proton nuclear elastic scattering
in FLUKA v4-3.4~\cite{angeli-csikai}. While the shapes match
(\textit{i.e.}\ the sampling scheme is effective), the intensity is not
correctly reproduced (\textit{i.e.}\ the FLUKA v4-3.4 integrated cross
section is often not sufficiently accurate). A self-consistent scheme
has been adopted instead, wherein the integrated cross section for the
nuclear elastic scattering of protons, $\sigma_\mathrm{(p,el)}$, has
been obtained by numerically integrating the fitted parametrized
expression \eqref{eq:fullexpression}. The red solid curve in
Fig.~\ref{fig:xs} displays the sampled DXS obtained with this
self-consistent approach, which is in perfect agreement with the fitted
DXS.

In FLUKA v4-4.0, special care has been devoted to ensure that at energies below Coulomb barrier, proton nuclear elastic scattering is inactive, and that only Coulomb scattering is accounted for at these energies. Thus, below Coulomb barrier, the integrated cross section for proton nuclear elastic scattering has been set to zero, and finite-size form factors in Coulomb scattering (see Section \ref{sec:ranft}) have been set to unity. In an energy window within $\pm5$\% of the Coulomb barrier, the integrated cross section for proton nuclear elastic scattering is gradually switched on, and conversely the finite-size form factors for Coulomb scattering are allowed to deviate from 1. Finally, at higher energies, both quantities take their full values.

\subsection{Error incurred when treating Coulomb and nuclear elastic scattering as separate interaction mechanisms}
\label{sec:disentangling}

As discussed in Section~\ref{sec:ranft}, FLUKA's use of the Moli\`ere MCS theory necessarily implies a separate treatment of Coulomb and nuclear elastic scattering which, as shown in Section~\ref{sec:partial-wave}, faces formal difficulties. In this section, the error incurred by such a scheme is assessed.

\begin{figure}[ht]
  \centering
  \begin{subfigure}[b]{\columnwidth}
    \centering
    \includegraphics[width=\columnwidth]{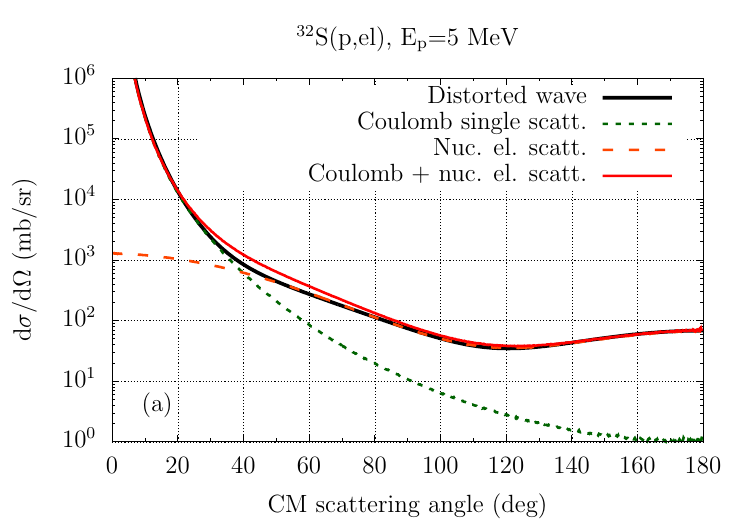}
  \end{subfigure}
  \hfill
  \begin{subfigure}[b]{\columnwidth}
    \centering
    \includegraphics[width=\columnwidth]{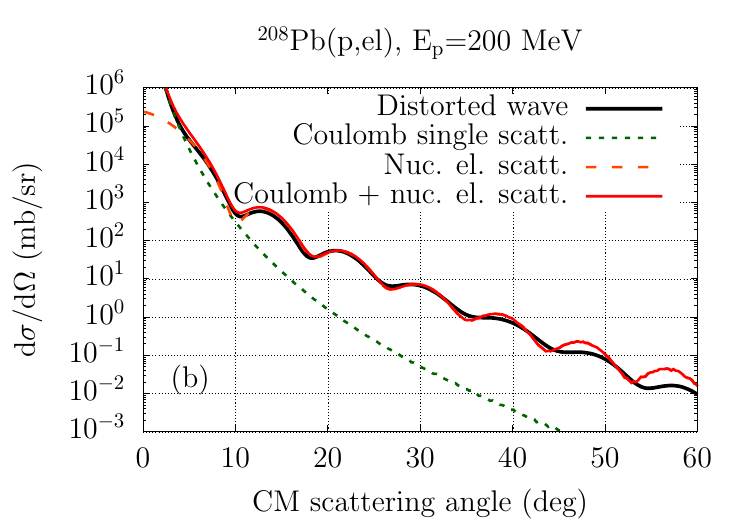}
  \end{subfigure}
  \caption{(a) Error incurred when treating Coulomb and nuclear elastic scattering as separate interaction mechanisms for 5~MeV protons on $^{32}$S. (b) Same as (a) for 200~MeV protons on $^{208}$Pb. See text for detailed discussion.}
  \label{fig:pricetopay}
\end{figure}

A dedicated benchmark has been carried out, whereby Coulomb and nuclear elastic scattering events have been sampled with FLUKA v4-4.0. The combined DXS resulting from the sum of Coulomb and nuclear elastic scattering events (each scaled with their respective integrated cross section) has been compared in absolute units of mb/sr with the corresponding experimental or partial-wave-calculated DXS for light or heavy targets, respectively. This comparison has been performed for a series of 27 target nuclei from $^{2}$H to $^{238}$U on a grid of 30 energies from Coulomb barrier up to 250~MeV. Figure \ref{fig:pricetopay}a displays the output of this benchmark for 5~MeV protons elastically scattered from $^{32}$S. The thick black curve represents the DXS calculated with the partial-wave scheme of Section~\ref{sec:partial-wave}, the dotted dark-green curve represents the angular distribution of Coulomb single scattering events sampled with FLUKA v4-4.0, the dashed orange curve represents the angular distribution for nuclear elastic scattering sampled with the model presented here, while the red curve is the sum of the last two curves. At these rather low energies (near Coulomb barrier), and especially for light and intermediate target nuclei, localized overestimations at intermediate scattering angles are observed, of at most a few 10\%. Instead, at large scattering angles (relevant for the radiation-to-electronics effects assessment in Section~\ref{sec:R2E}) the agreement is by construction optimal. Finally, for energies well above Coulomb barrier and especially for heavy target nuclei, the incurred error is negligible, as shown in Fig.~\ref{fig:pricetopay}b. Incidentally, this benchmark also probes the good performance of the integrated cross section scheme described in Section~\ref{sec:integrated_xs}.

\begin{figure*}[ht]
\centering
\includegraphics[]{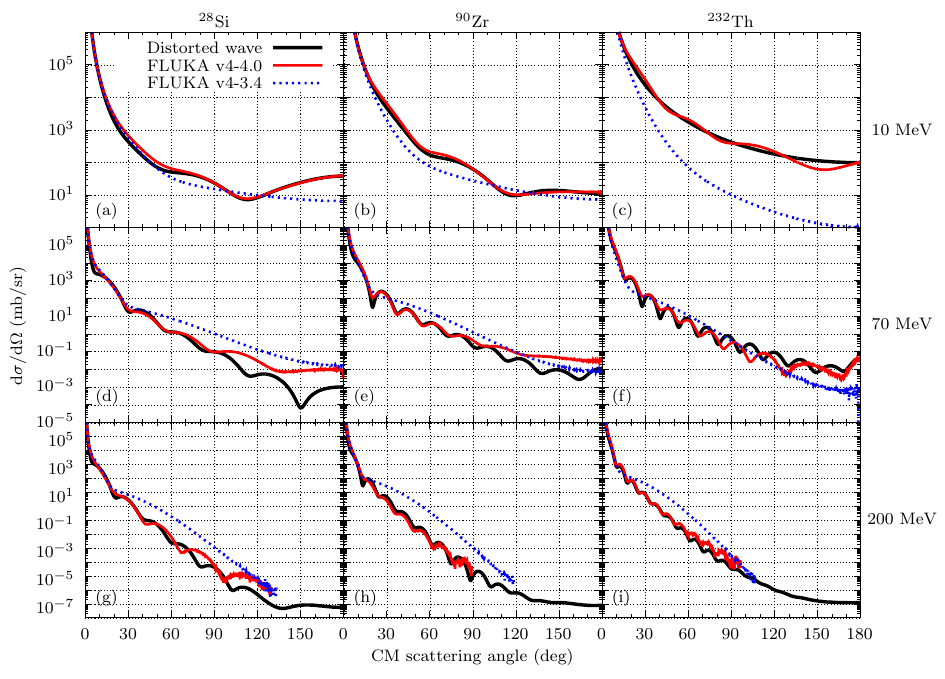}
\caption{Same as Fig.~\ref{fig:fits} for the full (Coulomb plus nuclear) elastic scattering of protons on nuclei. See text for detailed discussion.}
\label{fig:fits_Coulomb+nuc_ela}
\end{figure*}

To further gauge the error incurred by treating Coulomb and nuclear elastic scattering as separate interaction mechanisms, Fig.~\ref{fig:fits_Coulomb+nuc_ela} extends Fig.~\ref{fig:fits} with the inclusion of Coulomb elastic scattering, directly sampled from FLUKA v4-4.0 - with the relevant difference that now the thin red curves are sampled from (and not a direct evaluation of) parametrized expression \eqref{eq:fullexpression}. This comparison reveals the typically negligible error incurred by the separate treatment of Coulomb and nuclear elastic scattering, and an overall very good agreement, both in the forward scattering and in the large-scattering-angle domains. 

Additionally, Fig.~\ref{fig:fits_Coulomb+nuc_ela} shows that the proton nuclear elastic scattering model employed up to FLUKA v4-3.4 tends to either under- or over-estimate the actual DXS at large scattering angles. For heavy targets, for which the Coulomb barrier for protons approaches 10~MeV, the FLUKA v4-3.4 DXS for the elastic scattering (Coulomb plus nuclear) of protons with energies slightly above Coulomb barrier may underestimate the partial-wave DXS by orders of magnitude, as illustrated in subfigure (c). Instead, the FLUKA v4-4.0 model for proton nuclear elastic scattering presented here does not suffer from these artefacts.

\section{Application: SEU production in SRAMs under proton irradiation}
\label{sec:R2E}

As mentioned in Section~\ref{sec:intro}, earlier FLUKA simulations~\cite{coronetti,coronetti_paper} to estimate the cross section for SEU production in an ISSI SRAM~\cite{issi} under proton irradiation revealed discrepancies of up to two orders of magnitude against experimental cross sections in the 1-10~MeV proton energy range, see Fig.~\ref{fig:seu_intro}. This underestimation was mainly attributed to the lack of nuclear elastic scattering of protons below 10~MeV in FLUKA up to v4-3.4. In this section, the aforementioned simulations are revisited with FLUKA v4-4.0 in order to assess the performance of the newly developed model for proton nuclear elastic scattering presented in this work.

In Ref.~\cite{coronetti}, a further limitation of FLUKA impacting the simulation of SEU production in the ISSI SRAM under proton irradiation was pointed out. As of FLUKA v4-3.4, in the course of a Coulomb single scattering event, the direction of the charged projectile is updated, while its energy is not. The energy transferred to target atoms in the course of multiple Coulomb collisions is however accounted for in an average way along macroscopic particle steps via a nuclear stopping power term. This approach prevents event-by-event analyses, and therefore does not allow to assess the contribution of individual Coulomb collisions to the production of SEUs. Nevertheless, Coulomb collisions may contribute to SEU production, especially those with large scattering angle, suppressed as they may be~\cite{akkerman}. To quantify their event-by-event contribution to SEU production, a tentative closing of the Coulomb-single-scattering kinematics (as well as the explicit transport of the recoil where applicable) has been implemented in a development version of FLUKA. This model ingredient, amounting to an event-by-event account of nuclear-stopping-power effects, has been developed and tested for proton projectiles. An extension to heavier charged projectiles is underway and it is being intensively tested before being considered for public release in an ulterior version of FLUKA.

\begin{figure}
\centering
\includegraphics[width=\columnwidth]{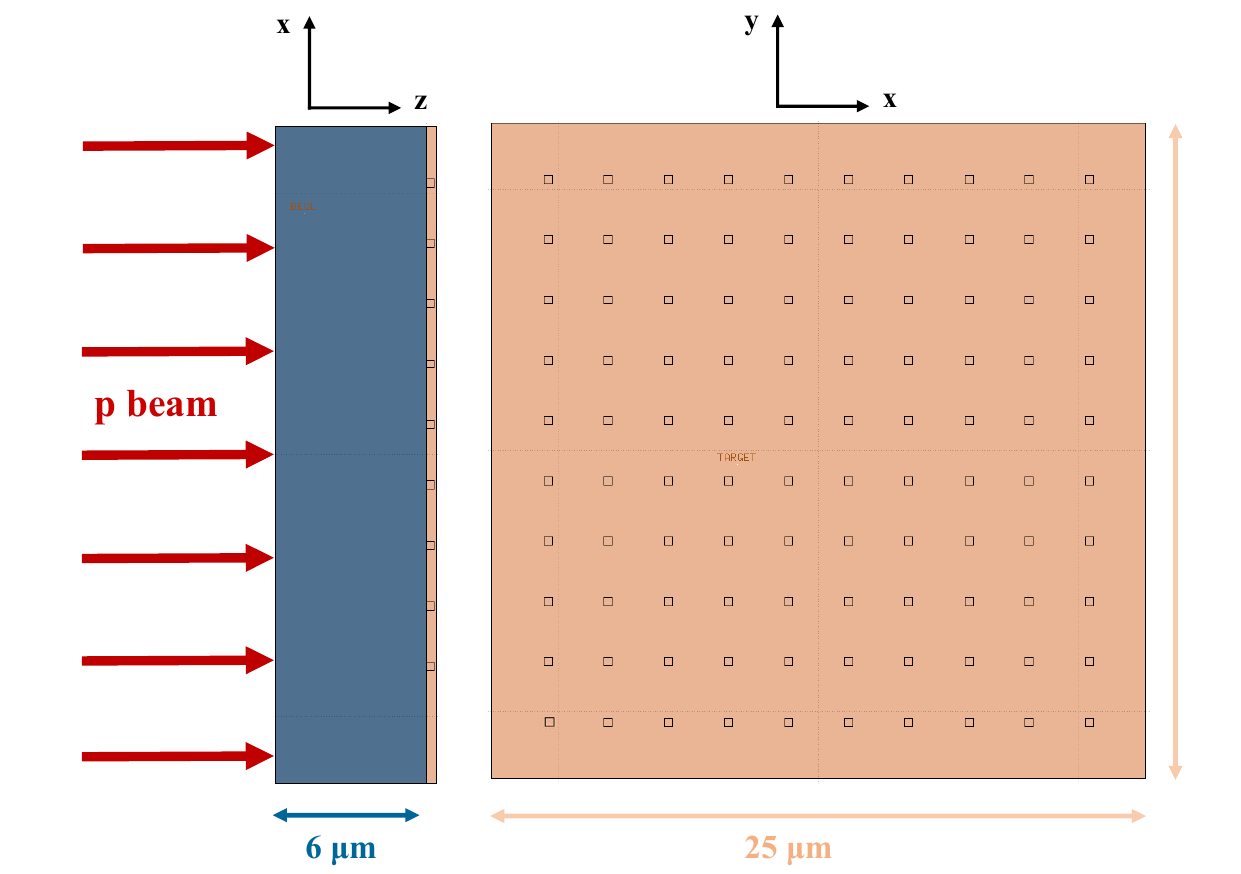}
\caption{ISSI SRAM simulation geometry in FLUKA.}
\label{fig:issi_geom}
\end{figure}

A simulation setup akin to that used in Ref.~\cite{coronetti} has been adopted. The sensitive volume (SV) of the ISSI SRAM is modelled in FLUKA as a 10~by~10 array of silicon cubes with a length of 310~nm embedded in a silicon matrix placed on a BEOL (back end of line) layer of SiO$_{2}$, as schematically displayed in Fig.~\ref{fig:issi_geom}. An extended parallel proton beam, covering the transverse face of the device, is considered, with energies ranging from 600~keV to 200~MeV. For each considered energy, the average energy deposition spectrum deposited in a cube, d$N/$d$E_\text{dep}$, has been scored on an event-by-event basis.

\begin{figure}
\centering
\includegraphics[width=\columnwidth]{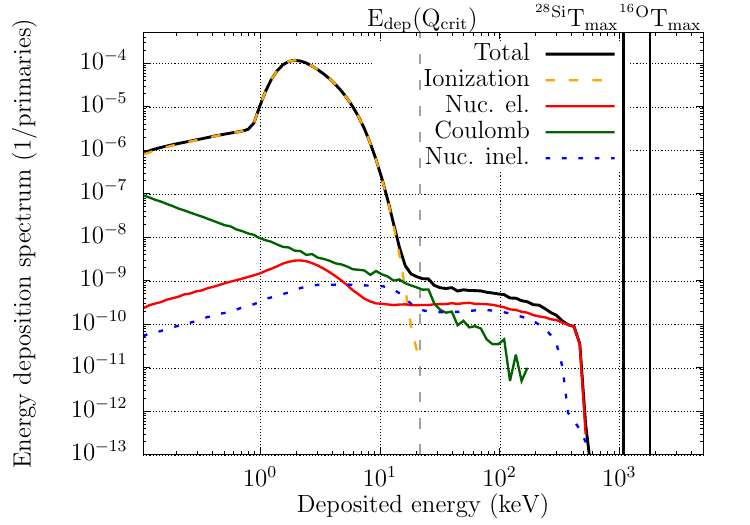}
\caption{Energy deposition histogram of 8~MeV protons in the ISSI SRAM, displaying the contributions from various kinds of particle histories. See text for detailed discussion.}
\label{fig:ene_hist}
\end{figure}

Figure \ref{fig:ene_hist} shows the simulated d$N/$d$E_\text{dep}$ for 8~MeV protons (thick black curve). The spectrum is further resolved into contributions of various kinds of particle histories, exploiting FLUKA's particle latching capabilities. The main peak (dashed yellow curve) corresponds to proton histories where there was only direct ionization by incident protons. The additional feature extending up to $\sim$600~keV is due to proton histories where a nuclear elastic scattering (red curve), a Coulomb single scattering (dark-green curve), or a nuclear reaction (dashed blue curve) occurred. For the first two kinds of histories, where a nuclear-elastic or a Coulomb collision occurred, energy is deposited in the SV by the recoiling target nucleus. The maximum kinetic energy transfer $(\mathrm{T}_\mathrm{max})$ during an elastic collision of a proton of 8~MeV on a target $^{28}$Si or $^{16}$O nucleus is of the order of 1~MeV and 1.8~MeV, respectively, displayed by the black vertical lines in Fig.~\ref{fig:ene_hist}. Note that the maximum recoil energy is rarely deposited entirely in the SV: some of the delta-rays (knock-on electrons) produced by the elastic recoil may leave the SV. Still, energies of the order of hundreds of keV (well beyond what protons typically impart by direct ionization) are easily deposited in the SV. Instead, for histories where a nuclear inelastic interaction occurred (dashed blue curve), energy is deposited in the SV by the residual nucleus or by emitted light fragments.

In order to trigger a SEU, a threshold energy deposition in the SV must take place. This value is typically given in terms of a threshold displaced charge (electron-hole pairs), the so-called critical charge $Q_\text{crit}$. For the ISSI SRAM, $Q_\text{crit}=~0.96$~fC~\cite{coronetti,coronetti_paper}, corresponding to a critical deposited energy $E_\text{dep}(Q_\text{crit})=~21.6$~keV, represented by the dashed grey line in Fig.~\ref{fig:ene_hist}. Note that the energy deposited in the SV by direct ionization does not exceed $E_\text{dep}(Q_\text{crit})$. Thus, for 8-MeV protons, direct ionization plays a negligible role in SEU production in the ISSI SRAM, which is instead driven by nuclear interactions. While energy deposition in the SV by Coulomb scattering events (dark-green curve) in this setup may reach $E_\text{dep}(Q_\text{crit})$, these kind of events are orders of magnitude less effective than nuclear elastic scattering (red curve) at producing higher energy depositions.

\begin{figure}
  \centering
  \includegraphics[width=\columnwidth]{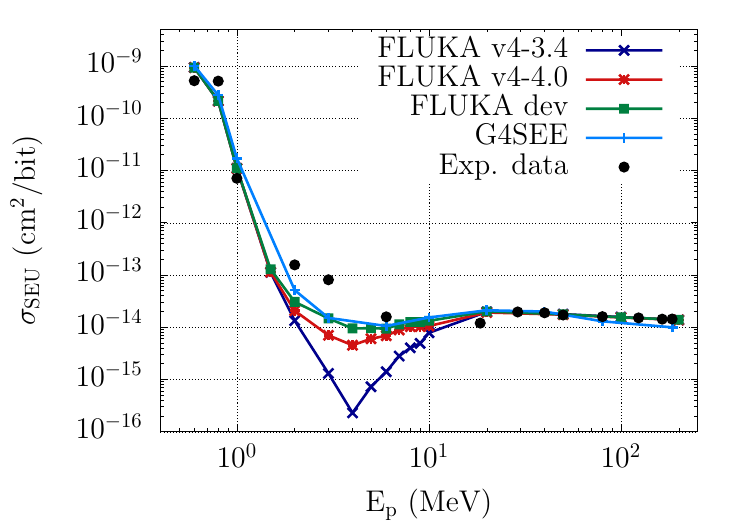}
  \caption{SEU production cross section in the ISSI SRAM as a function of the incident proton energy. See text for detailed discussion.}
  \label{fig:seu_final}
\end{figure}

The cross section for the production of SEUs, $\sigma_\mathrm{SEU}$, can be obtained for a given proton energy $E_\mathrm{p}$ as follows:
\begin{equation}
\label{eq:sigma_SEU}
\sigma_\text{SEU} (Q_\text{crit}; E_{\text{p}}) 
= 
\frac{1}{\Phi} 
\int \displaylimits_{E_\text{dep}(Q_\text{crit})}^{E_\text{max}}  \mathrm{d}E_\text{dep}\;\frac{\mathrm{d}N}{\mathrm{d}E_\text{dep}},
\end{equation}
where $\Phi$ is the incoming beam fluence. Figure~\ref{fig:seu_final} displays $\sigma_\mathrm{SEU}$ as a function of the proton energy.
Experimental values~\cite{coronetti,coronetti_paper} are displayed by the black dots with uncertainties smaller than the symbol size.
FLUKA v4-3.4 SEU production cross section estimates obtained with the prescription
above are shown in crosses (connected with solid line to guide the eye) and exhibit the aforementioned
underestimation of two orders of magnitude with respect to the
experimental cross section in the 1-10~MeV range. SEU production cross sections obtained with
FLUKA v4-4.0 (benefiting from the new model for proton nuclear elastic
scattering discussed in this work) are displayed by stars. Note that in
the energy range between 1-10~MeV the agreement with experimental cross
sections improves by a factor 20, although there are still local
underestimations. Thus, the newly developed model for the nuclear
elastic scattering of protons, included in FLUKA v4-4.0,  significantly
contributes to close the gap with respect to experimental SEU cross
sections. Finally, SEU production cross sections obtained with the FLUKA
development version (additionally including a tentative closing of
Coulomb-single-scattering kinematics together with an ad hoc biasing
scheme for nuclear elastic scattering) shown in squares, provide a
slightly better agreement with experimental cross sections in the
2-6~MeV region. 

\begin{figure}
  \centering
  \includegraphics[width=\columnwidth]{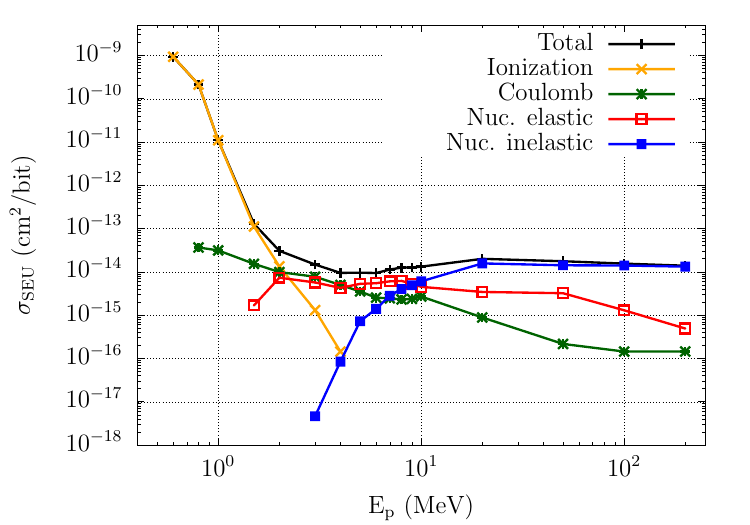}
  \caption{SEU production cross section in the ISSI SRAM induced by protons resolved into contributions of various kinds of particle histories. See text for detailed discussion.}
  \label{fig:seu_split}
\end{figure}

Similar results are obtained using other Monte Carlo tools, \textit{e.g.}  
G4SEE~\cite{g4see}, a Geant4-based~\cite{geant4} application, as shown in Fig.~\ref{fig:seu_final} by the curve with plus symbols. Note that residual discrepancies remain with
respect to the experimental cross sections, themselves subject to
uncertainty. These discrepancies may be due to uncertainties in the determination of $Q_\text{crit}$, 
to approximations and simplifications in the adopted simulation setup (the dimensions of 
the SV are not exactly known, the definition of the BEOL surrounding material does not reflect its actual complexity, etc.), or to model aspects such as the adopted Coulomb barrier prescription in FLUKA v4-4.0, which defines the energy at which proton nuclear elastic scattering opens. Overall, however, the agreement with experimental data is remarkable.

Finally, the particle-latching capabilities of FLUKA provide a practical means to elucidate which interaction mechanisms govern the production of SEUs in the considered ISSI SRAM in various proton energy domains. Figure~\ref{fig:seu_split} displays the contribution to $\sigma_\text{SEU}$ due to various kinds of particle histories. For proton energies below $\sim$2~MeV, direct ionization by incident protons largely dominates. In the energy range from roughly 2 to 8~MeV, recoils from nuclear elastic scattering (as well as Coulomb single scattering) events are the main contributors, while above $\sim$8~MeV contributions from recoiling residuals, as well as fragments from nuclear reactions drive SEU production.

\section{Conclusions and outlook}
\label{conclusions}

Earlier attempts to simulate with FLUKA the cross section for the production of SEUs in an ISSI SRAM under proton irradiation revealed an underestimation of up to two orders of magnitude in the 1-10~MeV energy range. Preliminary assessments attributed this underestimation predominantly to the lack of nuclear elastic scattering of protons below 10~MeV in FLUKA up to version 4-3.4.

To overcome this limitation, as well as a too crude treatment of large-angle nuclear elastic scattering (among other shortcomings) at intermediate energies in FLUKA v4-3.4, a new model for the nuclear elastic scattering of protons from Coulomb barrier up to 250~MeV on nuclei ranging from $^2$H to $^{238}$U has been developed and included in the public release of FLUKA v4-4.0. A combined approach relying on a partial-wave analysis and experimental angular distributions has been adopted, based on an effective parametrized expression which offers sufficient flexibility to adapt to the rich structure of maxima and minima of the actual differential cross section. The employed parametrization provides a systematic way of separating nuclear and Coulomb elastic scattering, in spite of the underlying formal difficulties of this approach, even at proton energies near Coulomb barrier. The error incurred by this separate treatment has been assessed and deemed minimal for the applications where nuclear elastic scattering plays a dominant role. The new model overcomes the unphysical lack of nuclear elastic scattering of protons below 10~MeV in previous versions of FLUKA and, furthermore, it provides a more realistic description of the structure of minima and maxima at large scattering angles typically present in the differential cross section for proton energies above a few tens of MeV.

The performance of this newly developed model for the nuclear elastic scattering of protons below 250~MeV has been assessed in the production of SEUs in an ISSI SRAM under proton irradiation. Whereas FLUKA v4-3.4 underestimated the SEU production cross section by up to two orders of magnitude for 1-10~MeV protons, the new model for proton nuclear elastic scattering in FLUKA v4-4.0 has led to an improved agreement by a factor of 20. A minor, but still important further improvement has been obtained by properly closing the Coulomb-single-scattering kinematics and explicitly generating the elastic recoil when applicable (instead of accounting for it globally via a nuclear-stopping-power term). This development feature is currently being tested and shall be made available in an ulterior public release of FLUKA, along with a scheme for the biasing of nuclear elastic scattering events.

The performances of FLUKA for a quantitative estimation of radiation effects in commercial electronic components have therefore significantly improved. Simulations presented here for the ISSI SRAM under proton irradiation are currently being extended to address SEU production in further commercial SRAMs and will be the subject of an upcoming benchmark work. In addition, nuclear elastic scattering of protons plays a relevant role in other applications. For instance, it governs the off-axis dose deposition by 50-250~MeV proton beams in water phantoms~\cite{verbeek,hall}, thus directly affecting dose delivered in hadron therapy to healthy tissue near cancer cells. The performances of the newly developed model for the nuclear elastic scattering of protons below 250~MeV have been recently assessed in the proton dosimetry context and will be documented in an upcoming publication.

Finally, the employed parametrized differential cross section, Eq.~\eqref{eq:fullexpression}, may be applicable to describe the nuclear elastic scattering of heavier ions (d, t, $^3$He, $^4$He, and beyond), missing as of FLUKA v4-4.0, where their elastic scattering is purely electrostatic. Moreover, Eq.~\eqref{eq:fullexpression} may also be applied to refine FLUKA's nuclear elastic scattering of neutrons above 20 MeV, relying on Ref.~\cite{ranft} as of FLUKA v4-4.0.

\bibliographystyle{elsarticle-num} 
\bibliography{ref}

\begin{thebibliography}{10}
\expandafter\ifx\csname url\endcsname\relax
  \def\url#1{\texttt{#1}}\fi
\expandafter\ifx\csname urlprefix\endcsname\relax\def\urlprefix{URL }\fi
\expandafter\ifx\csname href\endcsname\relax
  \def\href#1#2{#2} \def\path#1{#1}\fi

\bibitem{SEP}
G.~C. Messenger, M.~S. Ash, Single \uppercase{E}vent \uppercase{P}henomena
  \uppercase{I}, Springer US, 1997, Ch.~6, pp. 179--231.

\bibitem{avionics}
J.~Barak, N.~Yitzhak, \uppercase{SEU} \uppercase{R}ate in \uppercase{A}vionics:
  \uppercase{F}rom \uppercase{S}ea \uppercase{L}evel to \uppercase{H}igh
  \uppercase{A}ltitudes, IEEE Transact. Nuc. Sci. 62 (2015) 3369--3380.

\bibitem{huhtinen}
M.~Huhtinen, F.~Faccio, Computational method to estimate \uppercase{S}ingle
  \uppercase{E}vent \uppercase{U}pset rates in an accelerator environment,
  Nucl. Instrum. Meth. A 450~(1) (2000) 155--172.

\bibitem{ruben2017}
R.~G. Alía, \textit{et al.}, Single event effects in high-energy accelerators,
  Semicond. Sci. Technol. 32~(3) (2017) 034003.

\bibitem{sierawski}
B.~D. Sierawski, \textit{et al.}, Impact of \uppercase{L}ow-\uppercase{E}nergy
  \uppercase{P}roton \uppercase{I}nduced \uppercase{U}psets on \uppercase{T}est
  \uppercase{M}ethods and \uppercase{R}ate \uppercase{P}redictions, IEEE Trans.
  Nuc. Sci. 56~(6) (2009) 3085--3092.

\bibitem{cannon}
E.~H. Cannon, \textit{et al.}, Heavy \uppercase{I}on,
  \uppercase{H}igh-\uppercase{E}nergy, and \uppercase{L}ow-\uppercase{E}nergy
  \uppercase{P}roton \uppercase{SEE} \uppercase{S}ensitivity of 90-nm
  \uppercase{RHBD SRAM}s, IEEE Transact. Nuc. Sci. 57~(6) (2010) 3493--3499.

\bibitem{caron}
P.~Caron, \textit{et al.}, Physical \uppercase{M}echanisms of
  \uppercase{P}roton-\uppercase{I}nduced \uppercase{S}ingle-\uppercase{E}vent
  \uppercase{U}pset in \uppercase{I}ntegrated \uppercase{M}emory
  \uppercase{D}evices, IEEE Transact. Nuc. Sci. 66~(7) (2019) 1404--1409.

\bibitem{coronetti}
A.~Coronetti, Relevance and guidelines of radiation effect testing beyond the
  standards for electronic devices and systems used in space and at
  accelerators,
  \url{https://cds.cern.ch/record/2799812/files/CERN-THESIS-2021-255.pdf}
  (2021).

\bibitem{coronetti_paper}
A.~Coronetti, \textit{et al.}, Assessment of \uppercase{P}roton
  \uppercase{D}irect \uppercase{I}onization for the \uppercase{R}adiation
  \uppercase{H}ardness \uppercase{A}ssurance of \uppercase{D}eep
  \uppercase{S}ubmicron \uppercase{SRAM}s \uppercase{U}sed in \uppercase{S}pace
  \uppercase{A}pplications, IEEE Transact. Nuc. Sci. 68~(5) (2021) 937--948.

\bibitem{akkerman}
A.~Akkerman, J.~Barak, N.-M. Yitzhak, Role of \uppercase{E}lastic
  \uppercase{S}cattering of \uppercase{P}rotons, \uppercase{M}uons, and
  \uppercase{E}lectrons in \uppercase{I}nducing
  \uppercase{S}ingle-\uppercase{E}vent \uppercase{U}psets, IEEE Transact. Nuc.
  Sci. 64 (2017) 2648--2660.

\bibitem{zhenyu}
Z.~Wu, \textit{et al.}, Recoil-\uppercase{I}on-\uppercase{I}nduced
  \uppercase{S}ingle \uppercase{E}vent \uppercase{U}psets in
  \uppercase{N}anometer \uppercase{CMOS SRAM U}nder
  \uppercase{L}ow-\uppercase{E}nergy \uppercase{P}roton \uppercase{R}adiation,
  IEEE Transact. Nuc. Sci. 64~(1) (2017) 654--664.

\bibitem{flukaweb}
\url{https://fluka.cern}.

\bibitem{batt}
G.~Battistoni, \textit{et al.}, Overview of the \uppercase{FLUKA} code, Annals
  Nucl. Energy 82 (2015) 10--18.

\bibitem{frontiers}
C.~Ahdida, \textit{et al.}, New \uppercase{C}apabilities of the
  \uppercase{FLUKA} \uppercase{M}ulti-\uppercase{P}urpose \uppercase{C}ode,
  Front. in Phys. 9 (2022).

\bibitem{lerner}
G.~Lerner, \textit{et al.}, Analysis of the \uppercase{P}hotoneutron
  \uppercase{F}ield \uppercase{N}ear the \uppercase{TH}z \uppercase{D}ump of
  the \uppercase{CLEAR} \uppercase{A}ccelerator at \uppercase{CERN}
  \uppercase{W}ith \uppercase{SEU} \uppercase{M}easurements and
  \uppercase{S}imulations, IEEE Transact. Nuc. Sci. 69~(7) (2022) 1541--1548.

\bibitem{cecchetto}
M.~Cecchetto, \textit{et al.}, 0.1-10 \uppercase{M}e\uppercase{V}
  \uppercase{N}eutron \uppercase{S}oft \uppercase{E}rror \uppercase{R}ate in
  \uppercase{A}ccelerator and \uppercase{A}tmospheric \uppercase{E}nvironments,
  IEEE Transact. Nuc. Sci. 68~(5) (2021) 873--883.

\bibitem{ruben_thesis}
R.~G. Alía, Radiation fields in high energy accelerators and their impact on
  single event effects, \url{https://cds.cern.ch/record/2012360} (2014).

\bibitem{issi}
\url{https://www.issi.com/WW/pdf/61-64WV204816BLL.pdf} (2016).

\bibitem{ferrari}
A.~Ferrari, \textit{et al.}, An improved multiple scattering model for charged
  particle transport, Nucl. Instrum. Meth. B 71~(4) (1992) 412--426.

\bibitem{tsai}
Y.-S. Tsai, Pair production and bremsstrahlung of charged leptons, Rev. Mod.
  Phys. 46 (1974) 815--851.

\bibitem{ranft}
J.~Ranft, {Estimation of radiation problems around high-energy accelerators
  using calculations of the hadronic cascade in matter}, Part. Accel. 3 (1972)
  129--161.

\bibitem{exfor}
N.~Otuka, \textit{et al.}, Towards a \uppercase{M}ore \uppercase{C}omplete and
  \uppercase{A}ccurate \uppercase{E}xperimental \uppercase{N}uclear
  \uppercase{R}eaction \uppercase{D}ata \uppercase{L}ibrary
  \uppercase{(EXFOR)}: \uppercase{I}nternational \uppercase{C}ollaboration
  \uppercase{B}etween \uppercase{N}uclear \uppercase{R}eaction \uppercase{D}ata
  \uppercase{C}entres (\uppercase{NRDC}), Nucl. Data Sheets 120 (2014)
  272--276.

\bibitem{zerkin}
V.~Zerkin, B.~Pritychenko, The experimental nuclear reaction data
  (\uppercase{EXFOR}): \uppercase{E}xtended computer database and
  \uppercase{W}eb retrieval system, Nucl. Instrum. Meth. A 888 (2018) 31--43.

\bibitem{angeli-csikai}
J.~W. Wilson, \textit{et al.}, Nucleon-nucleus interaction data base-total
  nuclear and absorption cross sections, NASA TM-4053 (1988).

\bibitem{koning}
A.~J. Koning, J.~P. Delaroche, Local and global nucleon optical models from 1
  ke\uppercase{V} to 200 \uppercase{M}e\uppercase{V}, Nuc. Phys. A 713 (2003)
  231--310.

\bibitem{radial}
F.~Salvat, J.~M. Fernández-Varea, \uppercase{RADIAL}: A \uppercase{F}ortran
  subroutine package for the solution of the radial \uppercase{S}chrödinger
  and \uppercase{D}irac wave equations, Comp. Phys. Comm. 240 (2019) 165--177.

\bibitem{auerbach}
E.~H. Auerbach, S.~O. Moore, Calculations of \uppercase{I}nelastic
  \uppercase{S}cattering of \uppercase{N}eutrons by \uppercase{H}eavy
  \uppercase{N}uclei, Phys. Rev. 135 (1964) B895--B911.

\bibitem{ferrari1998}
A.~Ferrari, P.~Sala, The \uppercase{P}hysics of \uppercase{H}igh
  \uppercase{E}nergy \uppercase{R}eactions, in: Proceedings Workshop on Nuclear
  Reaction Data and Nuclear Reactor Physics, Design, and Safety, World
  Scientific, 1998, p. 424.

\bibitem{dremin}
I.~M. Dremin, Elastic scattering of hadrons, Physics-Uspekhi 56~(1) (2013) 3.

\bibitem{grichine}
V.~M. Grichine, Geant4 hadron elastic diffuse model, Comp. Phys. Comm. 181~(5)
  (2010) 921--927.

\bibitem{g4see}
D.~Lucsányi, \textit{et al.}, \uppercase{G4SEE}: A
  \uppercase{G}eant4-\uppercase{B}ased \uppercase{S}ingle \uppercase{E}vent
  \uppercase{E}ffect \uppercase{S}imulation \uppercase{T}oolkit and {I}ts
  \uppercase{V}alidation \uppercase{T}hrough \uppercase{M}onoenergetic
  \uppercase{N}eutron \uppercase{M}easurements, IEEE Trans. Nuc. Sci. 69~(3)
  (2022) 273--281.

\bibitem{geant4}
S.~Agostinelli, \textit{et al.}, {\uppercase{GEANT}4--a simulation toolkit},
  Nucl. Instrum. Meth. A 506 (2003) 250--303.

\bibitem{verbeek}
N.~Verbeek, \textit{et al.}, Single pencil beam benchmark of a module for
  \uppercase{M}onte \uppercase{C}arlo simulation of proton transport in the
  \uppercase{PENELOPE} code, Med. Phys. 48~(1) (2021).

\bibitem{hall}
D.~C. Hall, \textit{et al.}, Validation of nuclear models in \uppercase{G}eant4
  using the dose distribution of a 177 mev proton pencil beam, Phys. Med. Biol.
  61 (2015) N1--N10.

\end{thebibliography}

\biboptions{sort&compress}

\end{document}